\documentclass[preprint,12pt]{elsarticle}

\usepackage{amssymb}
\usepackage{amsmath}

\journal{arXiv}
\usepackage[utf8]{inputenc}
\usepackage{graphicx} 
\usepackage{amsfonts}
\usepackage{xcolor}
\usepackage{comment}
\usepackage{caption}
\usepackage{subcaption}
\usepackage{booktabs}
\usepackage{enumitem}

\newcommand{\ul}[1]{\underline{#1}}
\newcommand{\Christina}[1]{{\color{orange} #1}}

\begin{document}

\begin{frontmatter}

    \title{Operator learning regularization for macroscopic permeability prediction in dual-scale flow problem}
    \author[1]{Christina Runkel}
    \author[2,5]{Sinan Xiao}
    \author[3]{Nicolas Boull\'e\corref{cor1}}
    \author[4]{Yang Chen\corref{cor1}}

    \cortext[cor1]{Corresponding authors}

    \affiliation[1]{organization={Department of Applied Mathematics and Theoretical Physics, University of Cambridge},
        postcode={CB3 0WA},
        city={Cambridge},
        country={UK}}
    \affiliation[2]{organization={Department of Mathematical Sciences, University of Bath},
        city={Bath},
        postcode={BA2 7AY},
        country={UK}}

    \affiliation[3]{organization={Department of Mathematics, Imperial College London},
        city={London},
        postcode={SW7 2AZ},
        country={UK}}

    \affiliation[4]{organization={Department of Mechanical Engineering, University of Bath},
        city={Bath},
        postcode={BA2 2AY},
        country={UK}}

    \affiliation[5]{organization={Shien-Ming Wu School of Intelligent Engineering, South China University of Technology},
        city={Guangzhou},
        postcode={511442},
        country={China}}

    \date{\today}

    \begin{abstract}
        Liquid composites moulding is an important manufacturing technology for fibre reinforced composites, due to its cost-effectiveness. Challenges lie in the optimisation of the process due to the lack of understanding of key characteristic of textile fabrics – permeability. The problem of computing the permeability coefficient can be modelled as the well-known Stokes-Brinkman equation, which introduces a heterogeneous parameter $\beta$ distinguishing macropore regions and fibre-bundle regions. In the present work, we train a Fourier neural operator to learn the nonlinear map from the heterogeneous coefficient $\beta$ to the velocity field $u$, and recover the corresponding macroscopic permeability $K$. This is a challenging inverse problem since both the input and output fields span several order of magnitudes, we introduce different regularization techniques for the loss function and perform a quantitative comparison between them.
    \end{abstract}




\end{frontmatter}

\section{Introduction}
Liquid composites moulding is an important manufacturing technology for fibre reinforced composites, due to its cost-effectiveness. Challenges lie in the optimization of the process due to the lack of understanding of key characteristic of textile fabrics – permeability. The difficulty is mainly related to the nature of multiple lengths scales being involved in the flow kinematics across this type of porous media. At microscale, the resin flows between individual fibres, which has a typical length of micrometer; whereas, larger pores of the size of millimeters exist between the fiber bundles that are woven together, which leads to a clear fluid region at this commonly-called mesoscale. If one considers the problem with a unified length scale, at millimeter order of magnitude, the microscale flow needs to be described with Darcy’s law, and the mesoscale flow with Stokes equation. Then, the problem becomes a two-domain problem, for which an interfacial behavior between the Darcy region and the Stokes region has to be introduced. Previous studies \cite{brinkman1949calculation} have shown that a term may be necessary to be added to the Darcy’s law to incorporate the effect of such an interface. This leads to the well-known Brinkman equation \cite{brinkman1949permeability}. The Stokes-Brinkman equation can then be formulated to convert the two-domain problem into a single domain problem, simplifying the solution procedure, see e.g. \cite{hwang2010numerical, mezhoud2020computation}. The unified single domain problem introduces a heterogeneous parameter $\beta=0$ in clear-fluid regions (macropores) and $\beta=\mu k^{-1}$ in solid porous regions (fibre bundles), which allows the two domains to be automatically considered in the solution procedure, without the need of ad-hoc interface conditions.

Solving such a heterogeneous problem has been subject of several previous works using numerical techniques, such as Finite Element method \cite{hwang2010numerical, xie2008uniformly}, isogeometric analysis \cite{evans2013isogeometric}, lattice Boltzmann method \cite{martys2001improved} and fast Fourier transform method \cite{mezhoud2020computation, chen2023high}. In our previous work, a solution algorithm based on the Fast Fourier Transform (FFT) technique has been proposed, with an enhancement in convergence rate by Anderson’s acceleration \cite{chen2023high}. This method offers a capability of parallel computing, enabling large-scale simulations over high-resolution microstructures, which is essential for textile fabrics with complex fibrous architectures. Despite its state-of-the-art computational efficiency, the proposed method requires high-performance computing resources, which is not commonly available to industry users.

Deep learning has excelled in almost all fields of computational sciences over the past decade with the area of manufacturing not being an exception. Operator learning refers to a field of deep learning concerned with approximating an unknown operator which is often the solution operator of a partial differential equation (PDE) \cite{boulle2023mathematical, kovachki2024operator, kovachki2023neural}. Given data pairs $(f, u)$ with $f \in U$ and $u \in V$ from function spaces over a domain $\Omega \subset \mathbb{R}^d$, one aims to approximate a (nonlinear) operator $A : U \rightarrow V$ mapping $f$ to $A(f)=u$ by an approximation $\hat{A}$ such that $\hat{A}(f') \approx A(f')$ on both training and unseen data. In many operator learning scenarios, one is interested in learning a quantity of interest~\cite{huang2024operator}, like the macroscopic permeability coefficient in our case, in addition to the solution operator of the PDE (i.e., predicting the velocity field in our case).

\begin{figure}[htbp]
    \centering
    \includegraphics[width=0.6\linewidth]{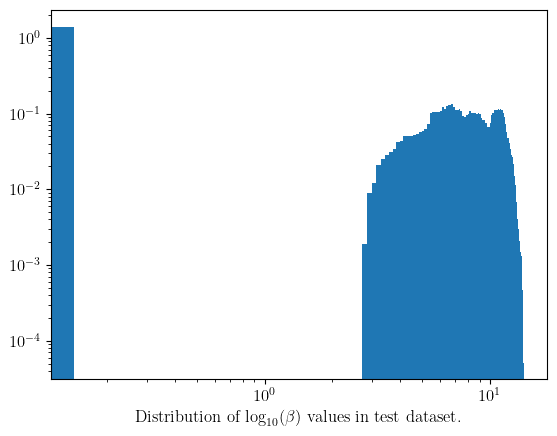}
    \caption{Distribution of heterogeneous coefficient $\beta$ in test dataset highlighting the scale of $\log_{10}\beta$. While many values of $\beta$ are zero and very close to zero, the coefficient can go up to values as large as $10^{14}$. The figure displays the distribution of magnitude values (in a logarithmic scale) on the training grid across the dataset considered.}
    \label{fig:distribution_beta_testset}
\end{figure}

This work aims to demonstrate the capability of operator learning to solving the challenging dual-scale flow problem, characterized by large variations of local permeability, hence involving extreme heterogeneity (spanning several orders of magnitude) in the governing equation ($\beta$ in the case of the unified Stokes-Brinkman equation). Fourier neural operators will be trained to learn the map from the heterogeneous coefficient $\beta$ to the macroscopic permeability coefficient $K$. As this is an inverse problem, computing $K$ from $\beta$ is challenging and computationally expensive when done with numerical solvers. One of the main challenges here lies in the large range of magnitudes of the parameter $\beta$ over which we want to predict the corresponding permeability coefficient. Figure \ref{fig:distribution_beta_testset} shows the distribution of the heterogeneous coefficient in the test dataset. While many values are equal to zero (clear-fluid region), the coefficient can be as high as $10^{14}$ for regions with very low permeability (due to densely packed fibers). We propose to approximate the solution in a two step process by first predicting a velocity field $u_{\text{pred}}$ that in the second step is then used to calculate the macroscopic permeability $K$.

The paper is organized as follows. We begin in section \ref{sec:fft_Stokes_Brinkman} with an overview of the Stokes--Brinkman problem and the FFT solver used for data generation. We then detail the data generation process in section \ref{sec:data_generation}, while section \ref{sec:no_Stokes_Brinkman} describes the neural operator learning approach to solve the Stokes-Brinkman problem. We highlight the different loss functions used for training the model before showing results for both the velocity field and the calculated macroscopic permeability in section \ref{sec:numerical_results}. Finally, we conclude and give an outlook on possible future work in section \ref{sec:conclusion}.

\section{Fast Fourier Transform solver for Stokes-Brinkman problem}\label{sec:fft_Stokes_Brinkman}
Considering a porous medium with two domains of distinct pore sizes: $\Omega = \Omega_f \cup \Omega_s$, the unified governing equation of the dual-scale flow is expressed in the form of Stokes-Brinkman equation:
\begin{equation}
    \label{eq:stokes_brinkman}
    \begin{aligned}
        \varphi \Delta u - \beta u - \nabla p & = 0, \quad \forall x \in \Omega, \\
        \mathrm{div}(u)                       & = 0, \quad \forall x \in \Omega, \\
    \end{aligned}
\end{equation}
with periodic velocity $u$ and pressure fluctuation $(p-G\cdot x)$ at $\partial \Omega$ as boundary conditions, and prescribed macroscopic velocity $\langle u \rangle_{\Omega}=U$ as loading condition. The heterogeneous coefficient functions $\varphi$ and $\beta$ are defined as
\begin{equation}
    \begin{split}
        \varphi(x) = \begin{cases}
                         \mu   & \text{if } x \in \Omega_f \\
                         \mu_e & \text{if } x \in \Omega_s
                     \end{cases}
    \end{split} \quad\text{and}\quad
    \begin{split}
        \beta(x) = \begin{cases}
                       0               & \text{if } x \in \Omega_f \\
                       \mu k_s^{-1}(x) & \text{if } x \in \Omega_s
                   \end{cases}
    \end{split},
\end{equation} where $k_s$ is the local permeability, which is assumed to be isotropic for simplicity. It is possible to apply the anisotropic local permeability in this solution framework as presented in \cite{chen2023high}. Here, $\mu_e =\mu$ is the so-called effective viscosity, which has been introduced to add a fictitious diffusion term to the Darcy’s flow in $\Omega_s$.

An FFT solver can be formulated by introducing two constant parameters $\varphi_0 = (\min \varphi + \max \varphi)/2$ and $\beta_0 = (\min \beta + \max \beta)/2$, which leads to a polarization vector $\tau = \big(\varphi- \varphi_0\big)\Delta u - \big(\beta-\beta_0\big) u$, allowing \eqref{eq:stokes_brinkman} to be re-arranged into
\begin{equation}
    \begin{aligned}
        \varphi_0 \Delta u - \beta_0 u - \nabla p + \tau = 0, & \quad \forall x \in \Omega, \\
        \mathrm{div}(u) = 0,                                  & \quad \forall x \in \Omega.
    \end{aligned}
\end{equation}
Considering $\tau$ as known at a given iteration, the above equation provides a closed form solution of the velocity field $u$ in the Fourier space as a function of $\tau$:
\begin{equation}
    \underline{\hat{u}} = \begin{cases}
        u,                                 & \text{if } \xi = 0    \\
        \hat{\mathbb{G}} \cdot \hat{\tau}, & \text{if } \xi \neq 0
    \end{cases},\quad \text{with }
    \hat{\mathbb{G}} = \frac{I -\frac{\xi\otimes \xi}{\|\xi\|^2}}{\varphi_0 \|\xi\|^2 + \beta_0},
\end{equation}
where $\hat{\mathbb{G}}$ is the Green operator, $u$ is the prescribed macroscopic velocity of the domain $\Omega$, and $\xi$ denote the frequency vector, whose definition is associated with the finite difference discretisation of the domain \cite{chen2023high}. Given that $\tau$ needs to be calculated from the velocity field at the previous iteration, a fixed-point algorithm can be thus formulated. Implementation details can be found in \cite{chen2023high}. In practice, the algorithm is enhanced with the Anderson acceleration \cite{walker2011anderson, chen2019analysis} to improve the convergence rate.

Once the velocity field $u$ is calculated with this FFT solver, a macroscopic pressure gradient can be derived as
\begin{equation}
    G = \langle \varphi \Delta u\rangle_{\Omega} - \langle \beta u\rangle_{\Omega},
\end{equation}
where $\langle \cdot \rangle_{\Omega}$ denotes for the volume average over the domain $\Omega$. The macroscopic permeability $\mathbf{K}$ of the bi-porous medium is then calculated by upscaling Darcy’s law:
\begin{equation}
    \label{eq:upscaled_darcys_law}
    U = - \frac{\mathbf{K}}{\mu} \cdot G.
\end{equation}
Due to the heterogeneity of the porous structure, the macroscopic permeability may be anisotropic, therefore, it is expressed by a $2 \times 2$ tensor $\mathbf{K}$ for the 2D setting considered in this study. In practice, to identify all the tensor components, we perform two independent simulations with loading conditions on the macroscopic velocity $U = [1,0]^T$ and $U = [0, 1]^T$, respectively.

\section{Data generation}\label{sec:data_generation}
We consider a 2D squared domain $\Omega$ of side length $1\text{mm}$, which is discretized into a uniform $N \times N$ structured grid. The viscosity parameters $\mu_e$ and $\mu$ are both set to $1 \text{MPa s}$. As mentioned in the previous section, two simulations are conducted using the velocity-controlled loading conditions of $U=[1, 0]^T \text{mm/s}$ and $U=[0,1]^T \text{mm/s}$, respectively. It should be noted that the fact that these parameters are fixed will not affect the proposed approach, because the end goal is to evaluate the macroscopic permeability, whose value is independent of the choice of the macroscopic velocity and viscosity (canceled out through Darcy’s law).

We generate 800 samples of input and output pairs $\{\beta_j, u_j\}$, with input being the heterogeneous coefficient $\beta$ and output being the corresponding velocity field $u$. We first sample the local permeability fields $\log (k_s)$ from a Gaussian random field, which are then used to generate samples of the heterogeneous coefficient $\beta$.

\begin{figure}[htbp]
    \centering
    \includegraphics[width=0.7\textwidth]{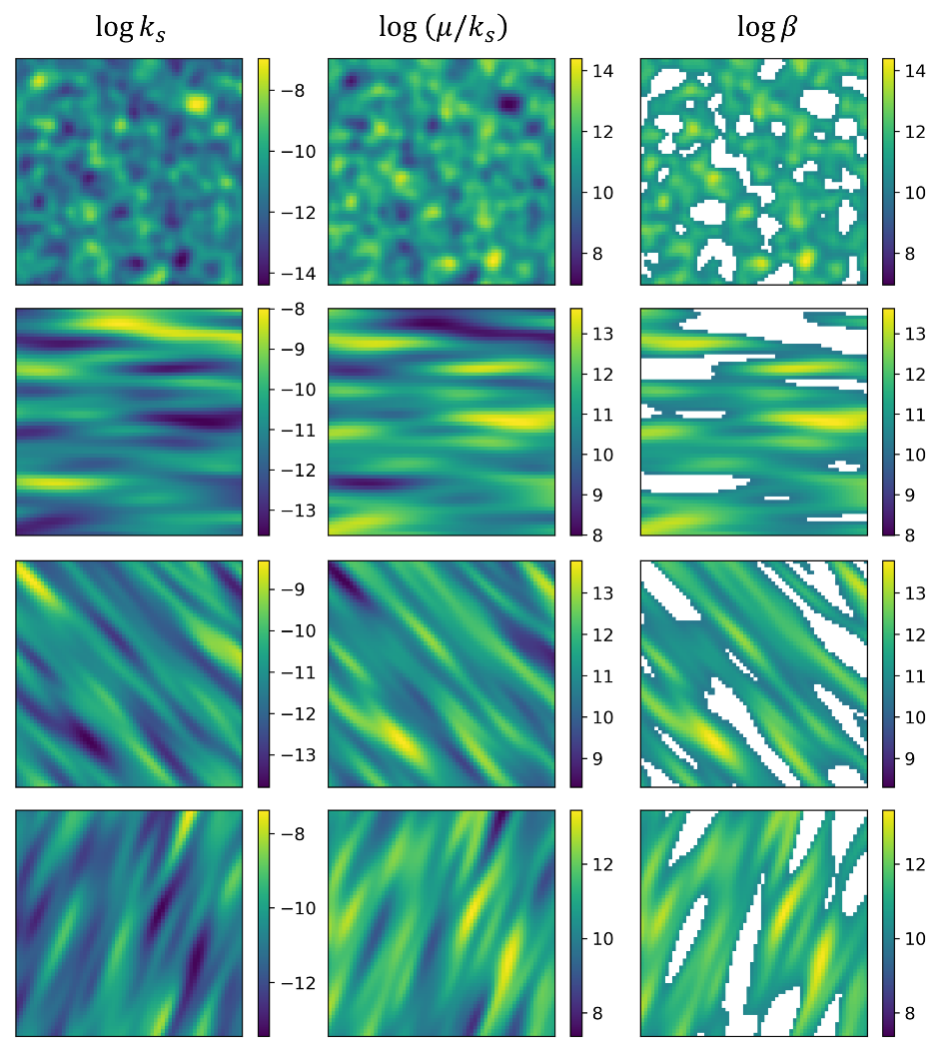}
    \caption{Examples of generated input: left – map of local permeability $k_s$; middle – map of $\mu/k_s$; right – map of heterogeneous coefficient $\beta$, the white region has value of $0$. All colour bars are shown in logarithm scale.}
    \label{fig:1}
\end{figure}

In this work, the fast Fourier transform moving average (FFT-MA) approach \cite{ravalec2000fft} is adopted for generating samples from a Gaussian random field (with a squared exponential kernel). Although this method is only suitable for regularly meshed grids, it is more computationally efficient than the Cholesky decomposition approach. For the FFT-MA approach, the covariance function is considered instead of the covariance matrix and the corresponding convolution product is determined efficiently in the frequency domain through FFT. We note that a generic yet computationally expensive way to generate samples from a Gaussian random field would be based on the Cholesky decomposition of the covariance matrix~\cite{williams2006gaussian}, allowing any grid structure and pixel location pattern to be handled.

With the samples of $\log (k_s)$, one can obtain the samples of $\log (\mu/k_s)$, as shown in Figure~\ref{fig:1}. This provides the map of the heterogeneous coefficient $\beta = \mu / k_s$ in the porous solid region $\Omega_s$ and $\beta = 0$ in the clear fluid region $\Omega_f$, with the latter being determined by thresholding the $80\%$ quantile of the map of $(\mu / k_s)$. We note that this threshold ($80\%$) could also be sampled differently, which would allow various volume fractions of clear-fluid region to be included. This is, however, not a focus of the present work. The varying parameters controlled in the data generation include
\begin{enumerate}
    \item Mean value of the local permeability: $[10^{-11}, 10^{-10}, \dots, 10^{-4}]\text{ mm}^2$.
    \item Scaling factor of the x-axis: $[0.05, 0.1625, 0.275, 0.3875, 0.5]$.
    \item Scaling factor of the y-axis: $[0.05, 0.1625, 0.275, 0.3875, 0.5]$.
    \item Rotation angle: $[0^{\circ}, 45^{\circ}, 90^{\circ}, 135^{\circ}]$.
\end{enumerate}

The wide range of local permeability ensures that the data is representative of various dual-scale porous structures, such as textile fabrics under different levels of compaction. The scaling factors control the anisotropy level along either the x- or y-axis: the greater the scaling factor, the more anisotropic the generated field becomes. This ensures that the data covers bi-porous media with different local structures, featuring either isotropically randomized local micropores or tubular micropores due to the alignment of continuous fibers. Additionally, the rotation angle is sampled to make the trained model orientation-insensitive.

We generate one sample for each set of parameters with the standard deviation fixed to a small value ($10^{-12}$). In total, 800 samples are generated, representing 800 sets of parameters. The field of view is 1mm $\times$ 1mm, and the data for the training stage has a resolution of 64 $\times$ 64 pixels. Unit cells with a higher resolution (128 $\times$ 128) are also generated to test the super-resolution capability of the trained model.
These input maps are fed into the numerical solver to produce the velocity field. The AMITEX version \cite{chen2023high, amitexwebsite} of the FFT solver was used.

\section{Neural operator learning for Stokes-Brinkman problem}\label{sec:no_Stokes_Brinkman}

We now introduce the operator learning technique used to predict the macroscopic permeability coefficient given an input heterogeneous coefficient $\beta$. Operator learning aims to approximate an operator by an analogue of a neural network, called a neural operator, that takes an input function $f: \Omega \rightarrow \mathbb{R}^{d_I}$ and outputs a function $u:\Omega \rightarrow \mathbb{R}^{d_O}$ on a compact domain $\Omega \subset \mathbb{R}^d$. In its most standard formulation~\cite{kovachki2023neural}, a neural operator is composed of a succession of integral operation layers followed by a pointwise application of a nonlinear activation function $\sigma$ as follows:
\begin{equation} \label{eq_int_no}
    u_{i+1}(x) = \sigma\Big(\int_{\Omega_i} K^{(i)}(x,y)u_i(y)dy +b_i(x)\Big), \quad x \in \Omega_{i+1}
\end{equation}
for a learnable integral kernel $K^{(i)}$, bias function $b_i$ at layer $i$. In recent years, several neural operator architectures have been proposed with the main neural operator architectures in the literature being DeepONet \cite{lu2021learning} and its extensions \cite{kontolati2023influence, lu2022comprehensive, kissas2022learning, oommen2022learning, de2022cost, jin2022mionet, goswami2022neural, zhu2019physics, wang2021learning, goswami2022physics}, graph neural operators like Graph Convolution Networks \cite{li2020neural} and Graph Attention Networks \cite{you2022nonlocal} as well as the Fourier Neural Operator \cite{li2020fourier} and its extensions \cite{tripura2022wavelet, you2022learning, li2024physics, konuk2021physics}.

In the case of the Fourier Neural Operator, the integral operator can be expressed as
\begin{equation}
    \int_{\Omega_i} K^{(i)}(x,y)u_i(y)dy = \mathcal{F}^{-1} ( \mathcal{R}_i \cdot \mathcal{F}(u_i))(x), \quad x \in \Omega_{i+1},
\end{equation}
where $\mathcal{R}$ denotes the complex multiplication in Fourier space by a weight matrix with subsequent truncation to $m$ Fourier modes, and $\mathcal{F}$ (resp. $\mathcal{F}^{-1}$) denotes the Fourier transform (resp. inverse Fourier transform). The output of the $i$-th layer can be then computed as
\begin{equation}
    u_{i+1} = \sigma \big(W_i u_i + \mathcal{F}^{-1} (\mathcal{R}_i \cdot \mathcal{F}(u_i)) + b_i \big),
\end{equation}
for a weight matrix $W_i$ acting on the features in the spacial dimension (note that this deviates slightly from \eqref{eq_int_no}). The activation function $\sigma$ is set to the Gaussian Error Linear Units (GELU) \cite{hendrycks2016gaussian}. This helps the operator to capture the second derivatives present in equation \eqref{eq:stokes_brinkman} that we consider. The overall model architecture is constructed by using a neural network $P$ to first lift the input function $f$ to a higher dimensional space and then applying the aforementioned Fourier layers recursively before projecting to the output dimension via a second neural network $Q$, i.e.,
\begin{equation}
    \Phi(f) := u_{\text{pred}} = Q \circ u_I \circ \dots \circ u_1 \circ P(f).
\end{equation}

The Tensorized Fourier Neural Operator (TFNO) \cite{kossaifi2023multi} is a more recent architecture that exploits Tucker factorization \cite{hitchcock1927expression} to factorize the weights of the neural operator. Tucker factorization approximates the full weight tensor $W$ by a low-rank factorization, i.e.,
\begin{equation}
    W \approx G \times_1 A^{(1)} \cdots \times_N A^{(N)},
\end{equation}
for a core tensor $G$, $N$ factor matrices $A^{(n)}$ and $\times_n$ being the mode-$n$ product.

Computing the macroscopic permeability $K$ from heterogeneous coefficients $\beta$ is extremely challenging and computationally intensive when done with numerical solvers. We thus make use of the advantages of operator learning to facilitate this problem solving process. As further detailed in \cite{huang2024operator}, the quantity of interest in many operator learning scenarios is not learning the solution operator of a PDE but rather a quantity of interest, the macroscopic permeability coefficient in our case. In a first step, we therefore attempted to learn the parameter to observable map by training a neural operator mapping from $\beta$ to $K$ directly. This however proved to be too unstable because of the inversion of the macroscopic pressure gradient $G$ during training. In a second step, we thus trained the neural operator with a relative $L^2$ loss function to enforce the predicted velocity field $u_{\text{pred}}$ to be close to the ground truth velocity field $u_{\text{GT}}$. While yielding sufficient results on the predicted velocity field $u_{\text{pred}}$, the computed macroscopic permeability differed drastically from the ground truth macroscopic permeability. We thus tried a $H^1$ loss function instead to enforce smoothness on the velocity field. In addition to that, we also propose and compare the following regularization strategies to address the issues mentioned above:
\begin{enumerate}[left=0pt]
    \item[a)] \textbf{Relative $L^2$ loss}: Enforcing the predicted velocity field to be close to the ground truth, we optimise using a relative $L^2$ loss function:
          \begin{equation}
              l_{L^2}(u_{\text{pred}}, u_{\text{GT}}) := \frac{\| u_{\text{pred}} - u_{\text{GT}} \|_{L^2(\Omega)}}{\| u_{\text{GT}}\|_{L^2(\Omega)}},
          \end{equation}
          for $L^2(\Omega)$ denoting the $L^2$ norm over the domain $\Omega$.
    \item[b)] \textbf{Relative $H^1$ loss}: To enforce smoothness on the velocity field, we introduce a relative $H^1$ loss function, i.e.,
          \begin{equation}
              l_{H^1}(u_{\text{pred}}, u_{\text{GT}}) := \frac{\| u_{\text{pred}} - u_{\text{GT}} \|_{H^1(\Omega)}}{\| u_{\text{GT}}\|_{H^1(\Omega)}},
          \end{equation}
          where
          \begin{equation}
              \| u \|_{H^1(\Omega)} = \Big(\|{u}\|_{L^2(\Omega)}^2 + \sum_{i=1}^d \|\frac{\partial {u}}{\partial x_i}\|_{L^2(\Omega)}^2\Big)^{1/2},
          \end{equation}
          and $L^2(\Omega)$ denoting the $L^2$ norm over the domain $\Omega$. The loss function is chosen to be the relative $H^1$ loss between the predicted velocity field and the ground truth (from the physics based FFT simulations).
    \item[c)] \textbf{$\mathbf{\beta u} $-regularized relative $H^1$ loss}: The aim is to compute the macroscopic permeability $K$ instead of just predicting the velocity field $u$. Since the heterogeneous coefficient $\beta$ has a large range in magnitude, this can lead to small differences between the output of the neural operator and the ground truth velocity field, yielding significant discrepancies in the calculation of the macroscopic permeability. Thus, we also test adding a $\beta u$ regularization term to penalize large differences between $\beta u_{\text{pred}}$ and $\beta u_{\text{GT}}$. This leads to:
          \begin{equation}
              l_{H^1 \beta u}(u_{\text{pred}}, u_{\text{GT}}) := \frac{\| u_{\text{pred}} - u_{\text{GT}} \|_{H^1(\Omega)}}{\| u_{\text{GT}}\|_{H^1(\Omega)}} + \lambda \frac{\| \beta u_{\text{GT}} - \beta u_{\text{pred}}\|_{L^1(\Omega)}} {\| \beta u_{\text{GT}}\|_{L^1(\Omega)}},
          \end{equation}
          where $L^1(\Omega)$ denoting the $L^1$ norm over the domain $\Omega$.

    \item[d)] \textbf{$\mathbf{\beta u} $ Laplacian-regularized relative $H^1$ loss}:
          The second term that can lead to discrepancies in the macroscopic permeability is the Laplacian of the velocity field $u$. In addition to the $\beta u$ regularization term, we therefore add a second term that penalizes for large differences between the Laplacian of $u_{\text{pred}}$ and $u_{\text{GT}}$. The resulting loss function thus reads:
          \begin{equation}
              \begin{aligned}
                  l_{H^1 \Delta \beta u}(u_{\text{pred}}, u_{\text{GT}}) & := \frac{\| u_{\text{pred}} - u_{\text{GT}} \|_{H^1(\Omega)}}{\| u_{\text{GT}}\|_{H^1(\Omega)}} + \lambda_1 \frac{\| \beta u_{\text{GT}} - \beta u_{\text{pred}}\|_{L^1(\Omega)}} {\| \beta u_{\text{GT}}\|_{L^1(\Omega)}} \\
                                                                         & + \lambda_2 \frac{\| \Delta u_{\text{pred}} - \Delta u_{\text{GT}}\|^2_{L^2(\Omega)}}{\| \Delta u_{\text{GT}}\|^2_{L^2(\Omega)}}.
              \end{aligned}
          \end{equation}

    \item[e)] \textbf{Relative $H^2$ loss}:
          Instead of adding a regularization term for the Laplacian of the velocity field, we also test a $H^2$ loss that explicitly penalizes for large differences in the second partial derivatives of $u_{\text{pred}}$ and $u_{\text{GT}}$.
          \begin{equation}
              l_{H^2}(u_{\text{pred}}, u_{\text{GT}}) := \frac{\| u_{\text{pred}} - u_{\text{GT}} \|_{H^2(\Omega)}}{\| u_{\text{GT}}\|_{H^2(\Omega)}},
          \end{equation}
          where
          \begin{equation}
              \| u \|_{H^2(\Omega)} = \Big(\|{u}\|_{L^2(\Omega)}^2 + \sum_{i=1}^d \|\frac{\partial {u}}{\partial x_i}\|_{L^2(\Omega)}^2 + \sum_{i=1}^d \sum_{j=1}^d \Big\|\frac{\partial {u}}{\partial x_i x_j}\|_{L^2(\Omega)}^2\Big)^{1/2}.
          \end{equation}

    \item[f)] \textbf{$\mathbf{\beta u} $-regularized relative $H^2$ loss}:
          Similar to training with the $\beta u $-regularized $H^1$ loss, we also try using a $\beta u$-regularized $H^2$ loss function, i.e.,:
          \begin{equation}
              l_{H^2 \beta u}(u_{\text{pred}}, u_{\text{GT}}) := \frac{\| u_{\text{pred}} - u_{\text{GT}} \|_{H^2(\Omega)}}{\| u_{\text{GT}}\|_{H^2(\Omega)}} + \lambda \frac{\| \beta u_{\text{GT}} - \beta u_{\text{pred}}\|_{L^1(\Omega)}} {\| \beta u_{\text{GT}}\|_{L^1(\Omega)}}.
          \end{equation}

    \item[g)] \textbf{Macroscopic pressure gradient loss}: As the aim is to compute the macroscopic permeability $K$ as exact as possible, we also try optimizing for the macroscopic pressure gradient $G$. $K$ is computed by computing the inverse of the macroscopic pressure gradient $G := \langle \varphi \Delta u \rangle_{\Omega} - \langle \beta u \rangle_{\Omega}$, divided it by $\mu$. We thus define the macroscopic pressure gradient loss as minimising the macroscopic pressure gradient $G$, i.e.,
          \[
              l_{\text{G}}(u_{\text{pred}}, u_{\text{GT}}) := \frac{\| (\langle \Delta u_{\text{pred}}\rangle_{\Omega} - \langle \beta u_{\text{pred}} \rangle_{\Omega} ) - (\langle \Delta u_{\text{GT}}\rangle_{\Omega} - \langle \beta u_{\text{GT}} \rangle_{\Omega} )\|_{L^1(\Omega)}}{\|\langle \Delta u_{\text{GT}}\rangle_{\Omega} - \langle \beta u_{\text{GT}} \rangle_{\Omega}\|_{L^1(\Omega)}}. \]
\end{enumerate}

In addition to testing different loss functions, we test both the vanilla FNO model \cite{li2020fourier} as well as an FNO version that uses Tucker factorization to reduce the number of parameters in the model (TFNO) \cite{kossaifi2023multi}. We implement both models using the open-source library \texttt{neuraloperator} \cite{li2020neural, kovachki2023neural}. Both neural operators in this study consist of $256$ lifting channels, $64$ projection channels and $4$ hidden Fourier layers. They are trained using the Adam optimizer \cite{kingma2014adam} with a step learning rate scheduler that multiplies the learning rate by $\gamma = 0.8$ every 40 epochs for a total of $1000$ epochs with a batch size of $16$, $700$ samples for training, and $100$ samples for testing.


The training was conducted using a uniform grid of size $64 \times 64$ to discretize the input data. For training, both loading conditions of $U=[1,0]^T \text{mm/s}$ and $U=[0,1]^T \text{mm/s}$ were included. Using the $700$ unique training samples, we augmented the data to the loading condition $U=[0,1] \text{mm/s}$ by simply transposing the input $\log \beta(x)$ to swap the x- and y- axes, as well as transposing the ground truth output, i.e., generating training pairs of the form $\big\{(\log \beta)^T, u_{\text{GT}}^T\}$.

\section{Results of numerical experiments}\label{sec:numerical_results}

In the following we compare the output of the neural operator, i.e., the velocity field $u$ for training the neural operator with the loss functions defined above. Secondly, we compute the macroscopic permeability $K$ and analyze the differences between the ground truth and the predicted macroscopic permeability coefficient.

\subsection{Comparison of velocity fields}
\begin{figure}[htbp]
    \centering
    \hspace*{0pt}
    \begin{subfigure}[b]{0.9\textwidth}
        \includegraphics[width=\textwidth]{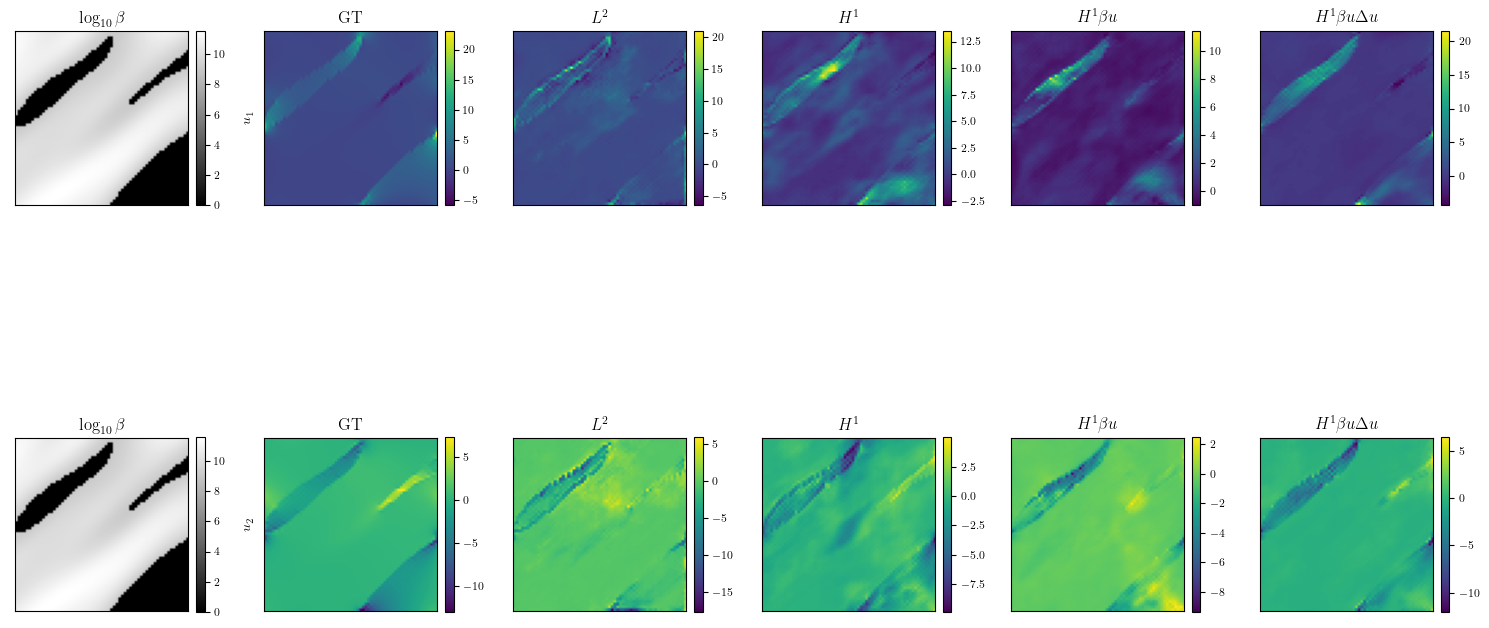}
    \end{subfigure}%
    \vspace{4em}
    \hspace*{0pt}
    \begin{subfigure}[b]{0.75\textwidth}
        \hspace*{0pt}
        \includegraphics[width=\textwidth]{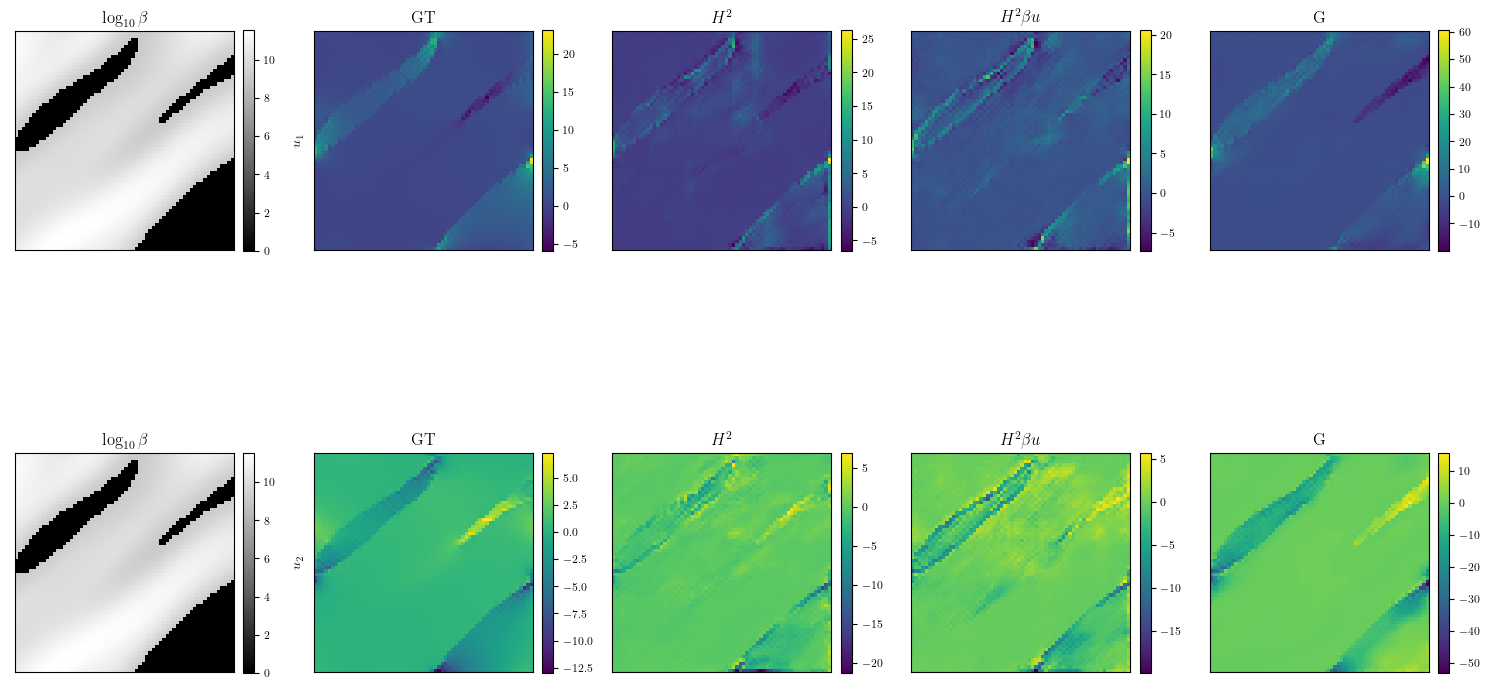}
    \end{subfigure}
    \caption{Comparison of ground truth (GT) \textbf{velocity field} $u_1$ and $u_2$ vs. the output of the \textbf{FNO model} trained on the seven different loss functions (top on and bottom plots) depicted in section \ref{sec:no_Stokes_Brinkman} for an example input $\log_{10} \beta$ of the test dataset. The best results for this example can be achieved for training with a $H^1 \beta u \Delta u$ loss function or when optimizing with a $H^1 \beta u$ loss. The neural operators that have been trained on the $H^1$ loss function and the macroscopic pressure gradient loss perform worst.}
    \label{fig:velocity_field_fno}
\end{figure}

\begin{figure}[htbp]
    \centering
    \hspace*{0pt}
    \begin{subfigure}[b]{0.9\textwidth}
        \includegraphics[width=\textwidth]{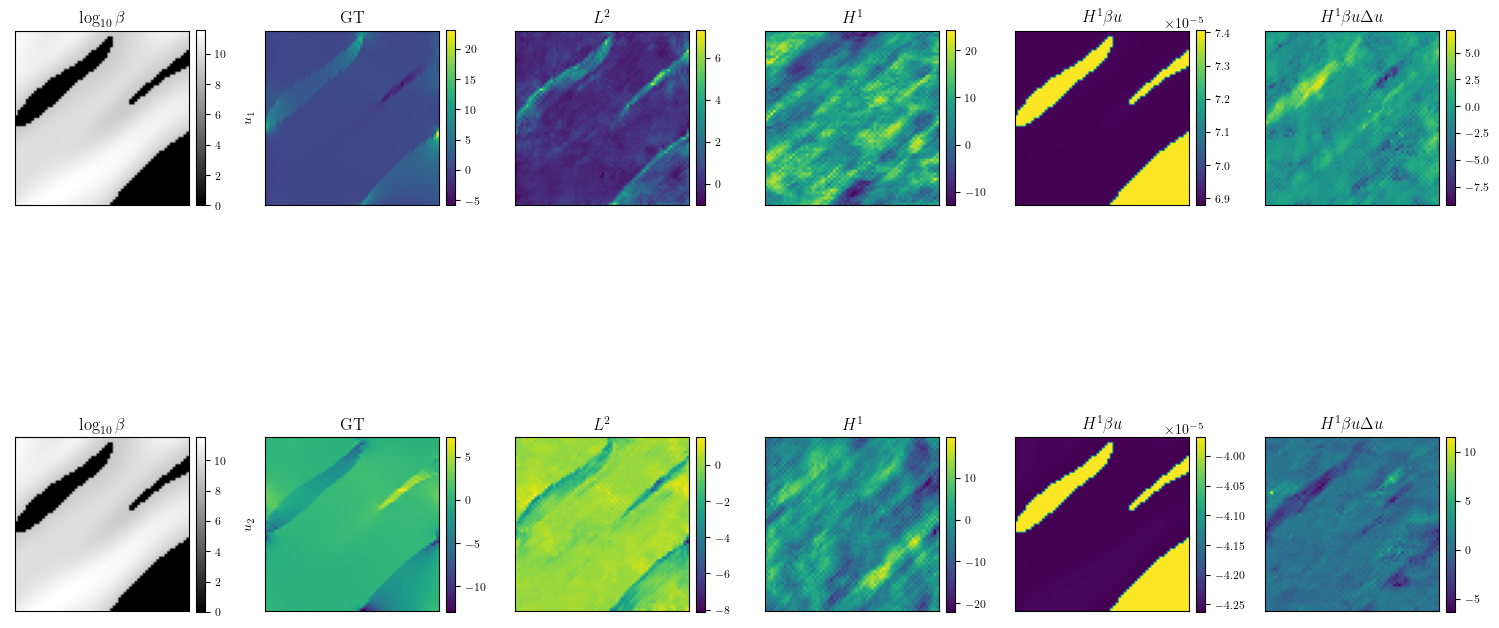}
    \end{subfigure}%
    \vspace{4em}
    \hspace*{0pt}
    \begin{subfigure}[b]{0.75\textwidth}
        \hspace*{0pt}
        \includegraphics[width=\textwidth]{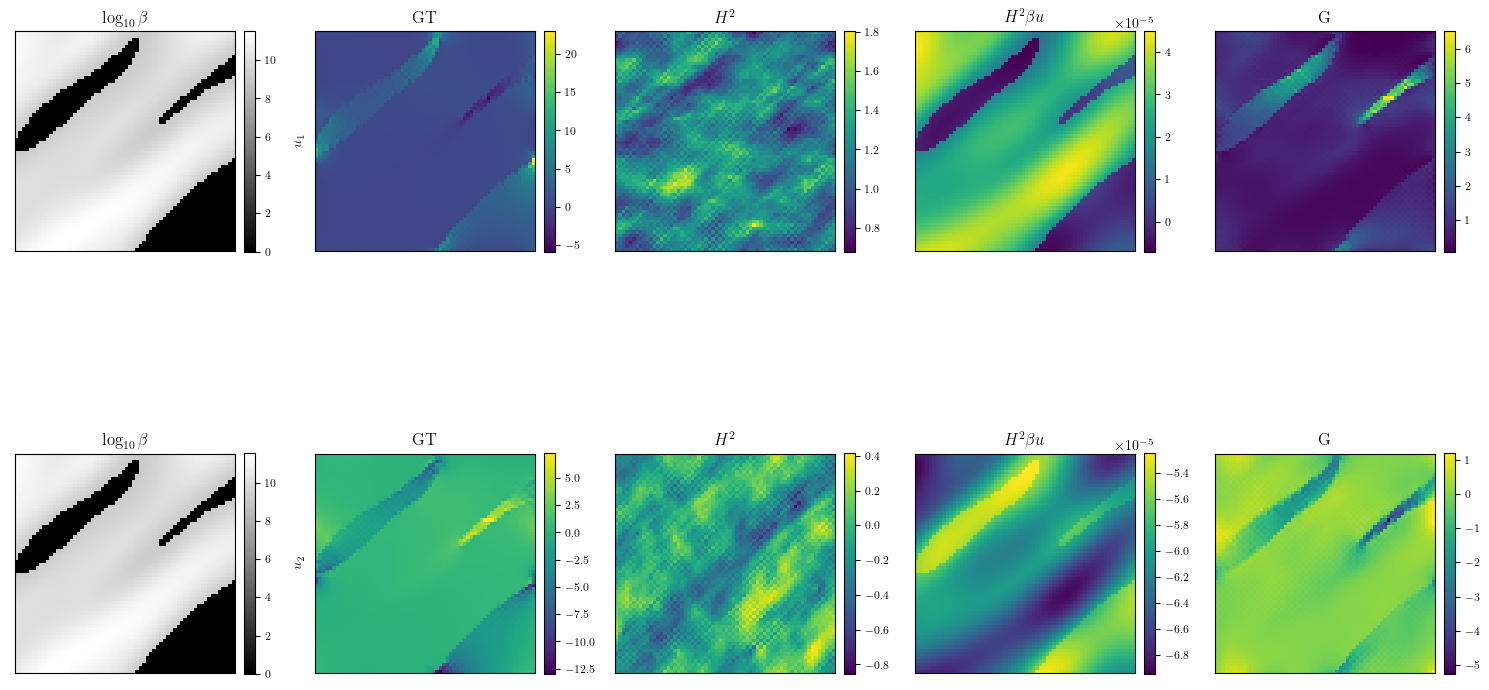}
    \end{subfigure}
    \caption{Comparison of ground truth (GT) \textbf{velocity field} $u_1$ and $u_2$ vs. the output of the \textbf{TFNO model} trained on the seven different loss functions depicted in Subsection \ref{sec:no_Stokes_Brinkman} for an example input $\log_{10} \beta$ of the test dataset. The best results for this example can be achieved for training with a $L^2$ loss function or when optimizing for the macroscopic pressure gradient. The neural operators that have been trained on the $H^1 \beta u$ and $H^2 \beta u$ loss function perform worst.}
    \label{fig:velocity_field_tfno}
\end{figure}

In a first step, we compare the output of the neural operator, i.e., the velocity field $u_{\text{pred}}$, to the ground truth output $u_{\text{GT}}$ computed via the FFT-MA method. Figures \ref{fig:velocity_field_fno} and \ref{fig:velocity_field_tfno} visually highlight the results for a randomly picked sample from the test dataset. The leftmost column shows the input $\log_{10}\beta$ with the second column from the left highlighting the ground truth images. Comparing the ground truth images to the outputs of the neural operator trained with the $L^2$, $H^1$, $H^1 \beta u$-regularized, $H^1 \beta u \Delta u$-regularized, $H^2$, $H^2 \beta u$-regularized and $G$ loss functions, training the FNO model with $H^1 \beta u \Delta u$ loss function and optimising for $H^1$ with an additional $\beta u$ regularisation term yields the best results. The neural operators trained with a $H^1$ loss function and the macroscopic pressure gradient perform worst.

For the TFNO model, Figure \ref{fig:velocity_field_tfno} highlights the visual results for an example from the training set. The best results can be achieved when training with a $L^2$ loss function or when optimizing for the macroscopic pressure gradient. The neural operators that have been trained on the $H^1 \beta u$ and $H^2 \beta u$ loss function perform worst.

\begin{table}[tbp]
    \centering
    \caption{Overview of normalized squared $L_2$ distance between ground truth \textbf{velocity field} $u_{\text{GT}}$ and output of the neural operator $u_{\text{pred}}$ trained with different loss functions on the test dataset. We normalize the squared $L_2$ distance by the norm of $u_{\text{GT}}$, i.e., $\|u_{\text{GT}\|_2^2}$. For the FNO model, training with a $H^1 \beta u$ loss yields the best results (highlighted in bold), followed by training with a $H^2 \beta u$ loss function (highlighted in italic). In case of the TFNO, the model trained on the macroscopic pressure gradient performs best (highlighted in bold and underlined) while the model trained on the $H^2$ loss function comes second (highlighted in italics). The TFNO model trained on the macroscopic pressure gradient loss performs best out of all models.}
    \begin{tabular}{c c c c c c c c} \toprule
        $L_2^2$ ($\downarrow$) & $L^2$    & $H^1$    & $H^1 \beta u$ & $H^1 \beta u \Delta u$ & $H^2$    & $H^2 \beta u$ & $G$      \\ \midrule
        \textbf{FNO}           & $1.1492$ & $1.3061$ & $0.9887$      & $1.6825$               & $1.2039$ & $1.1367$      & $3.5042$ \\
        \textbf{TFNO}          & $0.8180$ & $3.7899$ & $1.0000$      & $1.0867$               & $0.8812$ & $1.0000$      & $0.7098$ \\ \bottomrule%
    \end{tabular}
    \label{tab:normalized_l2_diff_loss_funcs}
\end{table}

The visual results are supported by Table \ref{tab:normalized_l2_diff_loss_funcs} which shows the relative squared $L_2$ distance between the ground truth velocity field and the output of the neural operators trained on the different loss functions computed on the whole test dataset. The metric was chosen to facilitate comparability between solutions and loss functions. For the FNO model, optimizing with an $H^1 \beta u$ loss function yields the best results while the outputs of the neural operators trained with the $H^2 \beta u$ loss function perform second best. For the TFNO model, the model trained on the macroscopic pressure gradient loss function performs best with the model trained on an $L^2$ loss function coming second. The TFNO model trained on the macroscopic pressure gradient loss outperforms all other models. While the models were trained successfully to achieve around $5-10\%$ relative error on the training set, capturing the right velocity scale for the velocity on the testing set is highly challenging as can be seen in Table \ref{tab:normalized_l2_diff_loss_funcs} and Figures \ref{fig:velocity_field_fno} and \ref{fig:velocity_field_tfno}. Despite these issues, we will see in the next section that one can still recover the macroscopic permeability scale accurately

\subsection{Comparison of macroscopic permeability}

\begin{figure}[htbp]
    \centering
    \includegraphics[width=0.8\textwidth]{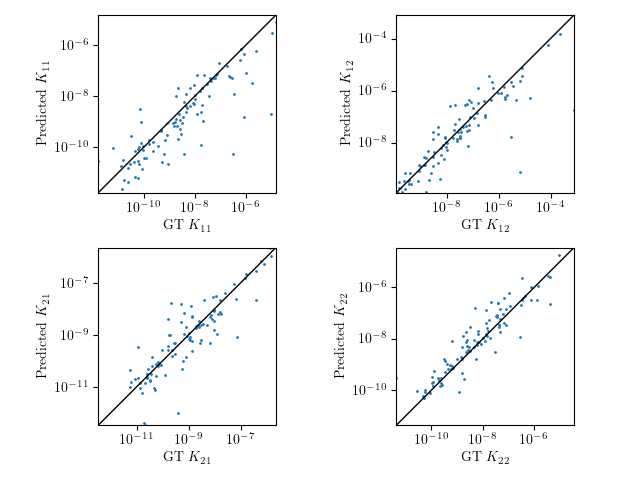}
    \caption{Ground truth \textbf{macroscopic permeability} $K_{\text{GT}}$ (x-axis) vs. predicted $K_{\text{pred}}$ (y-axis) for the best performing model. Training the TFNO model with an $H^2$ loss function yields the best results out of all the models and regularizations tested. Except for a few outliers, the computed macroscopic permeability for the test samples is close to the diagonal, implying that $K_{\text{pred}}$ is close to $K_{\text{GT}}$.}
    \label{fig:K_best_result}
\end{figure}

\begin{figure}[htbp]
    \centering
    \begin{subfigure}[b]{0.3\textwidth}
        \centering
        \includegraphics[width=\textwidth]{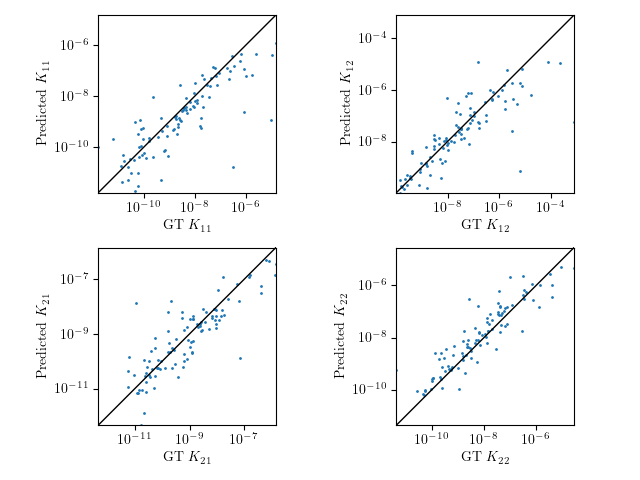}
        \caption{TFNO model trained on $H^1$ loss function with $\beta u \Delta u$-regularization term.}
        \label{fig:tfno_h§}
    \end{subfigure}
    \hfill
    \begin{subfigure}[b]{0.3\textwidth}
        \centering
        \includegraphics[width=\textwidth]{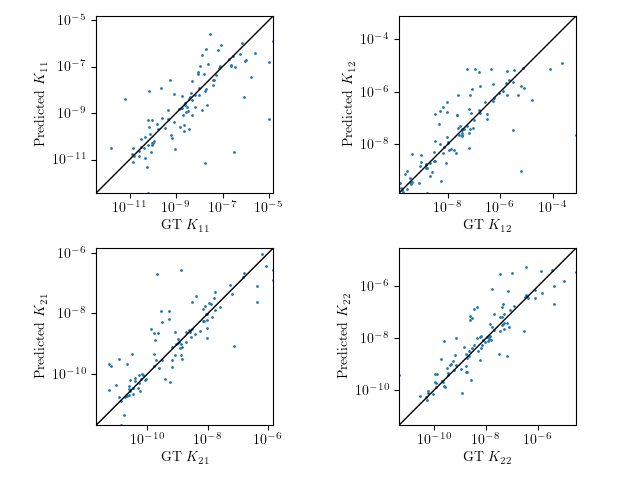}
        \caption{TFNO model trained on $L^2$ loss function. \\ \;}
        \label{fig:tfno_h1betau}
    \end{subfigure}
    \hfill
    \begin{subfigure}[b]{0.3\textwidth}
        \centering
        \includegraphics[width=\textwidth]{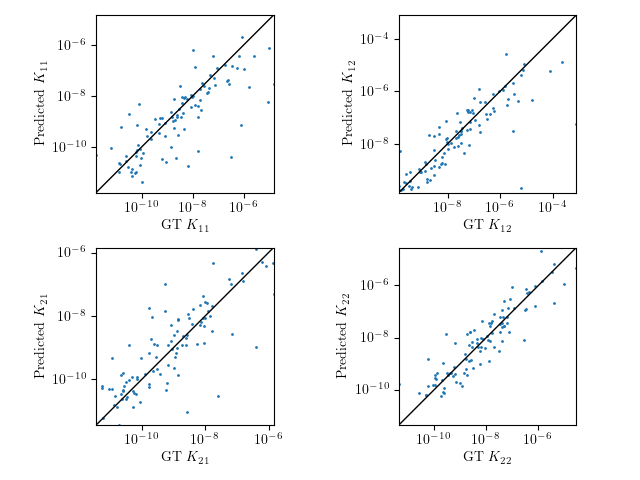}
        \caption{TFNO model trained on macroscopic pressure gradient loss function.}
        \label{fig:tfno_h1beatulaplacian}
    \end{subfigure}
    \caption{Ground truth \textbf{macroscopic permeability} $K_{\text{GT}}$ (x-axis) vs. predicted $K_{\text{pred}}$ (y-axis) for the models performing second to fourth best. Training the TFNO model with the $H^1$ loss with $\beta u \Delta u$-regularization term yields the second best results. Using the $L^2$ loss function to train the TFNO model produces the third best results whereas the TFNO model trained on a macroscopic pressure gradient loss function comes in fourth place.}
    \label{fig:K_second_to_fourth_best_model}
\end{figure}

Since we our primary motivation is to recover the macroscopic permeability rather than the velocity field, we additionally compare the differences between the macroscopic permeability computed with the ground truth velocity field $u_{\text{GT}}$ and those using the output of the neural operators. Figures \ref{fig:K_best_result} and \ref{fig:K_second_to_fourth_best_model} give an overview of the best result and the second to fourth best result, respectively.

The difference between ground truth and predicted macroscopic permeability $K$ for the FNO model is smallest when training the neural operator with an $H^2$ loss. The second best results are achieved when optimizing for the macroscopic pressure gradient loss function.

For the TFNO model, training with an $H^2$ loss yields best results while minimizing for an $H^1$ loss with $\beta u \Delta u$-regularization achieves the second best results in terms of the difference between $K_{\text{GT}}$ and $K_{\text{pred}}$. This highlights the importance of regularising for second order derivatives as we need to compute the Laplacian of the velocity field.

\begin{table}[htbp]
    \centering
    \caption{Comparison of ground truth \textbf{macroscopic permeability} $K_{\text{GT}}$ and predicted macroscopic permeability $K_{\text{pred}}$ for different types of loss functions and regularizations. We report the relative surpremum norm, i.e., $J := \frac{\|\log (K_{\text{pred}}) - \log (K_{\text{GT}})\|} {\|\log (K_{\text{GT}})\|}$ for $\|K\| = \{ \max_{i,j} |k_{ij}| \}$. The TFNO model outperforms significantly, with the $H^2$ loss function achieving the best results (highlighted in bold and underlined). The second and third best performing models are the TFNO models trained on a $H^1 \beta u \Delta u$ loss (highlighted in bold) and $L^2$ loss (highlighted in italics), respectively.}
    \begin{tabular}{c c c c c c c c} \toprule
        $J$ ($\downarrow$) & $L^2$             & $H^1$    & $H^1 \beta u$ & $H^1 \beta u \Delta u$ & $H^2$                  & $H^2 \beta u$ & $G$      \\ \midrule
        \textbf{FNO}       & $0.1811$          & $0.2011$ & $0.1877$      & $0.2030$               & $0.1719$               & $0.1896$      & $0.1757$ \\
        \textbf{TFNO}      & $\mathit{0.1018}$ & $0.1139$ & $0.6593$      & $\mathbf{0.0979}$      & $\ul{\mathbf{0.0913}}$ & $0.5343$      & $0.1098$ \\ \bottomrule%
    \end{tabular}
    \label{tab:K_diff_loss_funcs}
\end{table}

\subsection{Zero-shot super-resolution}
\begin{figure}
    \centering
    \includegraphics[width=0.8\linewidth]{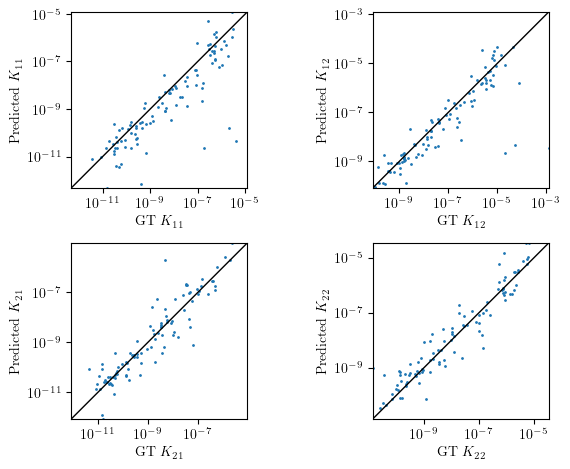}
    \caption{Ground truth macroscopic permeability $K_{GT}$ (x-axis) vs. predicted $K_{\text{pred}}$ (y-axis) for TFNO model and \textbf{zero-shot super-resolution}. We train the model with an $H^2$ loss on $64 \times 64$ grid, while testing on $128 \times 128$ grid. The values for the macroscopic permeability coefficient even for the higher discretisation are close to the diagonal, implying a good approximation from the neural operator.}
    \label{fig:tfno_SR}
\end{figure}

The Fourier Neural Operator and Tensorized Fourier Neural Operator are resolution independent models. This allows us to perform zero-shot super-resolution, i.e., train on a more corse grained grid while testing on a more fine grained grid. We thus perform experiments evaluating the best performing neural operator from the previous experiments that was trained on a $64 \times 64$ grid, i.e., the TFNO model trained on a $H^2$ loss function, on a $128 \times 128$ grid. Visual results for the macroscopic permeability coefficient on the test dataset are depicted in Figure \ref{fig:tfno_SR}. Even when testing on a more fine grained grid than the neural operator was trained on, the results for the macroscopic permeability resemble those for the $64 \times 64$ grid with the supremum norm being only slightly higher ($0.0971$ for $128 \times 128$ grid versus $0.0913$ for $64 \times 64$ grid).

\section{Conclusion}\label{sec:conclusion}
In the present work, we were interested in learning the map from the heterogeneous coefficient $\beta$ to the macroscopic permeability coefficient $K$. As computing this map is challenging and computationally expensive when done with numerical solvers, we proposed to replace the physics-based solver by a neural operator. We approximate the solution in a two step process by first predicting a velocity field $u_{\text{pred}}$ that in the second step is then used to calculate the macroscopic permeability $K$.

We showed that regularising the second derivatives of the velocity field during training significantly enhances the results for the computed macroscopic permeability as the computation includes computing the Laplacian of the velocity field $u$.
Training the neural operator with a $H^2$ or $H^1$ loss with additional $\beta u \Delta u$-regularization term yields a macroscopic permeability that is close to the ground truth solution computed by the numerical solver -- while having an almost negligible computational cost. In addition to that, we showed that as our approach is using the TFNO model, we are able to achieve good results in a zero-shot super-resolution setting.

To the best of our knowledge, this work presents the first application of operator learning in addressing the challenging dual-scale flow problem, characterized by significant variations in local permeability. Although the study utilized 2D examples, the findings are expected to be generalizable to 3D, which will necessitate greater computational resources for model training. Notably, only 700 samples were used to train our 2D model, yet the results are promising, especially in the case of zero-shot super-resolution tests. This data-efficient feature of FNO will be particularly advantageous when extending the method to 3D problems with more realistic microstructures of textile reinforcements.

\section*{Acknowledgements}
C.R. acknowledges support from the Cantab Capital Institute for the Mathematics of Information (CCIMI) and the EPSRC grant EP/W524141/1. This work was supported by the Office of Naval Research (ONR) under grant N00014-23-1-2729, and the Engineering and Physical Sciences Research Council (grant number EP/P006701/1), as part of the EPSRC Future Composites Manufacturing Research Hub.

\bibliographystyle{elsarticle-num}
\bibliography{biblio}
\end{document}